\documentclass[11pt]{article}

\usepackage[preprint]{acl}

\usepackage{times}
\usepackage{latexsym}
\usepackage[T1]{fontenc}
\usepackage[utf8]{inputenc}
\usepackage{microtype}
\usepackage{inconsolata}

\usepackage{graphicx}
\usepackage{subcaption}
\usepackage{booktabs}
\usepackage{float}          %
\usepackage{amsmath}
\usepackage{amssymb}
\usepackage{mathtools}
\usepackage{amsthm}
\usepackage[capitalize,noabbrev]{cleveref}

\theoremstyle{plain}

\theoremstyle{definition}

\theoremstyle{remark}

\usepackage[disable,textsize=tiny]{todonotes}

\title{GitScholar: A Dataset for Predicting AI Research Impact from GitHub Engagement}

\author{
  \textbf{Emilien Guandalino\textsuperscript{1,*}},
  \textbf{Lorenz K. M\"uller\textsuperscript{1}},
  \textbf{Beatrice Alessandra Motetti\textsuperscript{1,2}},
\\
  \textbf{Konstantin Berestizshevsky\textsuperscript{1}},
  \textbf{Lukas Cavigelli\textsuperscript{1}}
\\
  \textsuperscript{1}Computing Systems Lab, Huawei Research, Switzerland
\\
  \textsuperscript{2}Politecnico di Torino, Italy
}

\begin{document}
\maketitle
{\makeatletter\renewcommand\@makefntext[1]{\noindent#1}\makeatother\let\thefootnote\relax\footnotetext{\fontsize{8.5pt}{10pt}\selectfont\textsuperscript{*}\textbf{Correspondence:} \mbox{\href{mailto:emilien.baptiste.guandalino1@h-partners.com}{emilien.baptiste.guandalino1@h-partners.com}}}}

\begin{abstract}
   With the rapid pace of AI research and the hundreds of daily new publications, staying up-to-date with the latest developments has become increasingly difficult. For researchers, quickly identifying impactful work is essential, yet manually reviewing each new publication is impractical. Automated impact prediction methods help address this challenge, usually by combining various information sources available, such as a paper's content or citation history. In this work, we propose using GitHub engagement as an additional source and demonstrate that it provides both a timely and accurate signal. To this end, we introduce GitScholar, a novel dataset that links GitHub activity from 444,000 repositories to over 558,000 AI arXiv papers. Our experiments show that GitHub reactions improve early prediction precision by up to 12\% over a strong academic baseline. Additionally, we find that GitHub signal offers near-complete coverage of high-impact AI papers, and consistently correlates with future academic success. GitScholar is publicly available at \url{https://huggingface.co/datasets/huawei-csl/GitScholar}.
\end{abstract}

\section{Introduction}

The last ten years have marked a period of rapid progress in artificial intelligence (AI). Recent breakthroughs, particularly in generative AI, have delivered impressive results and attracted widespread global attention. This is reflected in the recent explosion of scientific publications emerging both from academia and industry. For instance, in 2019 an average of 5.26 new researchers entered the field every hour \cite{tang2020}, and between 2019 and 2023, the AI annual publication rates increased at a compound annual growth rate of 25.93\% \cite{abanga2024bibliometric}. As a consequence, AI has become increasingly competitive and early detection of emerging trends is essential to capitalize on new opportunities.

In addition, peer-reviewed journals and conferences, which have traditionally served to promote quality research, still rely on lengthy review processes that are misaligned with today's accelerated research cycles. Many researchers now release preprints on platforms such as arXiv to quickly disseminate their work, claim ownership of new ideas, and begin accumulating citations sooner~\cite{feldman2018,lariviere2014arxiv}. These preprints should, in theory, enable earlier detection of influential ideas compared to traditional academic channels. In practice, however, manually assessing the quality of unreviewed papers requires reading at least parts of them, which is impractical given the hundreds of new submissions each day. 

To address this challenge, automated impact prediction methods have been developed to estimate a paper's potential impact, typically measured by its future citation count. These methods draw on various academic information sources related to a paper  \cite{info8030073}. Some of these sources are static and available at the time of publication, such as the paper’s content, the author’s reputation, or the affiliated institution. Others accumulate over time, including citation count, venue acceptance, test-of-time awards, and others. However, there is often a gap between the immediate availability of static information and the delayed accumulation of dynamic metrics like citation count. During this gap, which can span several weeks to months, no further relevant information is available in the academic world.

This has prompted researchers to explore non-academic sources, for example social media platforms where academic discourse occurs \cite{eysenbach2011tweets}, such as $\mathbb{X}$ or Reddit. While this information is potentially noisier, it is available quickly, from days to a few weeks after publication, and can serve as a useful signal for a paper's immediate reception. However, data from major social media platforms is not freely accessible and can be costly, placing large-scale analyses beyond the reach of many academic researchers and smaller institutions.

In this paper, we propose using GitHub engagement as an additional source of signal, and investigate its suitability to assess the relevance of recent AI publications on arXiv. Intuitively, there are several reasons to pursue this direction. GitHub is a widely adopted collaborative platform within computer science, including the AI community \cite{michael2020}. Unlike non-technical platforms such as $\mathbb{X}$, where reactions reflect opinions from a broad public, GitHub interactions (e.g. stars, forks, issues, or pull requests) indicate practical interest from a specialized audience. Additionally, GitHub activity revolves around code and collaboration, rather than popularity. Most importantly, GitHub data is freely accessible through its public API, making it a viable resource for our use case.

Based on these intuitions, we curate a large-scale dataset centered around interactions between GitHub and AI arXiv papers, and evaluate its effectiveness for early-stage impact prediction. More formally, we make the following contributions:
\vspace{-0.5em}

\begin{itemize}
	\item {\bf Release the GitScholar dataset:} A large-scale, temporally annotated dataset capturing the relationship between GitHub engagement and the subsequent academic impact of AI arXiv papers. The temporal annotations allow users to reconstruct the dataset's state at arbitrary points in the past, enabling consistent backtesting. The dataset is released under FAIR (Findable, Accessible, Interoperable, and Reusable) data principles \cite{Wilkinson2016FAIR} to support transparency and reuse by the research community.
	\item {\bf Evaluation of GitHub data as an early signal:}
	      We perform experiments demonstrating that GitHub data enhances the precision of predicting future influential AI papers, especially within the initial weeks post-publication. Our findings show that GitHub covers nearly all high-impact AI papers, confirming its central role in the AI research ecosystem. Compared to a public snapshot of Twitter data, GitHub provides a stronger and cleaner predictive signal and has shown substantial improvement since.
\end{itemize}
\vspace{-0.5em}

Taken together, our findings demonstrate that GitHub data provides timely and reliable signal to assess the relevance of emerging AI papers. By releasing this dataset, we aim to foster further research in this area.

\section{The GitScholar Dataset}

\subsection{Data Description}

\begin{figure}[t]
	\centering
	\includegraphics[width=1\columnwidth]{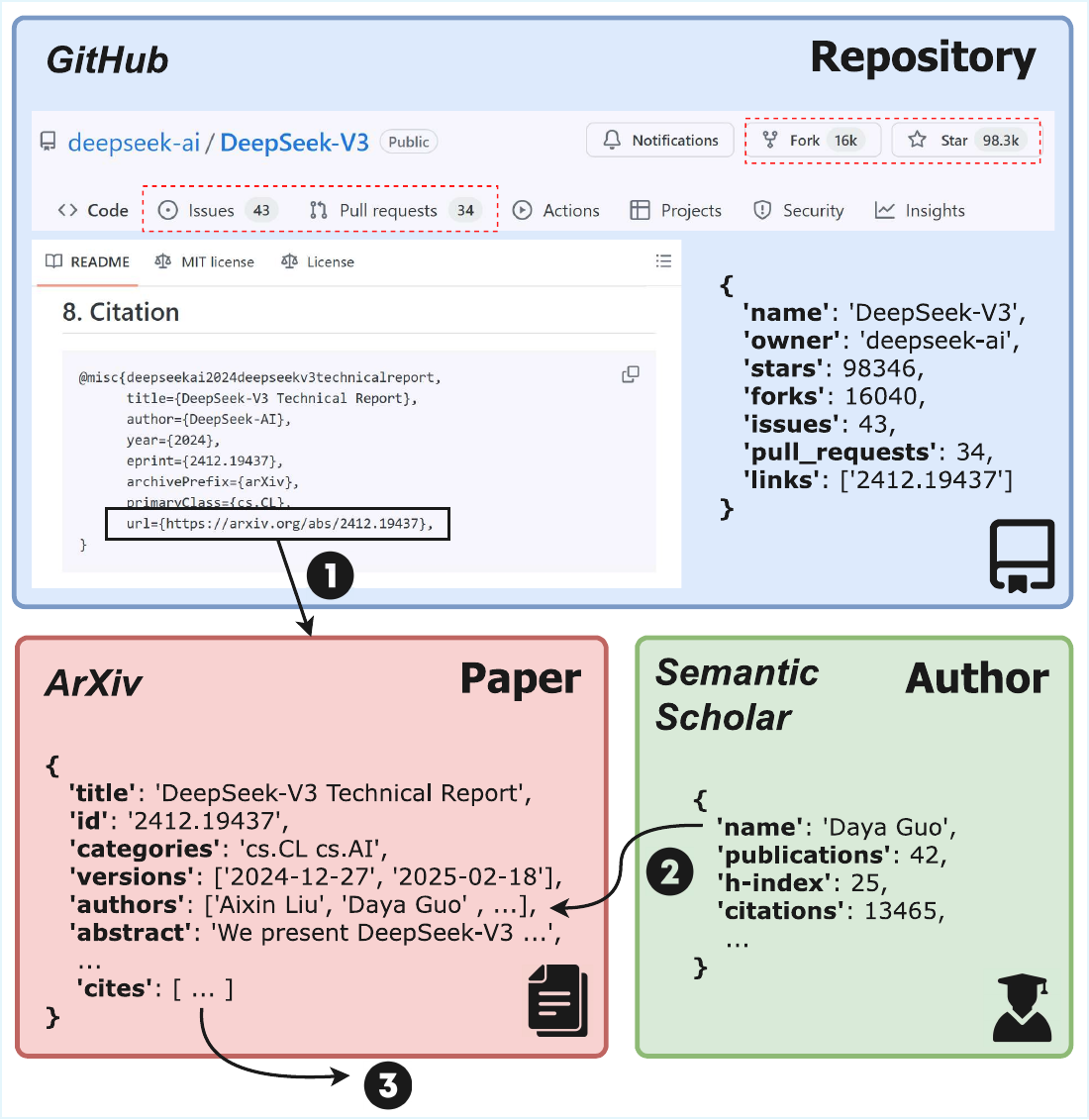}
	\caption{Overview diagram of the graph. The relationship edges are: (1) \emph{Repository-to-Paper}, i.e. mentions, (2) \emph{Author-to-Paper}, i.e. authorship, and (3) \emph{Paper-to-Paper}, i.e. citations. All features are timestamped (not shown here).}
	\label{fig:dataset}
\end{figure}

At a high level, our dataset is structured as a large heterogeneous graph that integrates academic metadata and GitHub activity. The nodes represent academic or GitHub entities, and the edges express the various relationships between them. An overview diagram is shown in Figure \ref{fig:dataset}.

\textbf{Academic Entities} There are two primary types of academic entities: \emph{papers} and \emph{authors}. The academic edges are either \emph{Paper-to-Paper}, representing a citation, or \emph{Author-to-Paper}, representing authorship. Each paper node contains standard metadata (e.g. title, abstract, arXiv category, submission history...), and each author node includes traditional academic indicators (e.g. $h$-index, $i$-10 index, total citation count....), computed from the sub-graph of their publications.

While our dataset focuses on papers available on arXiv, it includes citation relationships to and from papers outside arXiv. This ensures that citation-based metrics and author-level features are computed over the broader academic ecosystem, and not limited to the arXiv subset.

\textbf{GitHub Entities}
The GitHub entities in our graph correspond to \emph{code repositories}. The features of a repository node are its engagement metrics, i.e. its number of stars, forks, issues, and pull requests.
Our graph includes a \emph{Repository-to-Paper} edge for every arXiv paper that is mentioned in a repository's README file. These connections form the bridge between the academic and GitHub entities.

\textbf{Feature Timestamping}
To capture temporal dynamics, all entities and relationships are associated with a timestamp. More specifically, each GitHub repository has a recorded creation date and its features (star, fork, issue, pull request, arXiv mention) are timestamped based on when the events occurred. Each paper has a publication date, which determines the timestamps of outgoing citations and authorship edges. This temporal information enables users to filter the graph at any given point in time and reconstruct its state as of a specific date. Such granularity supports backtesting approaches, where models can be trained on historical data to predict future outcomes. Additionally, it allows for efficient, incremental updates of the dataset by querying only new or modified elements.

\textbf{Feature Extensibility} Each entity in the dataset is uniquely identified by one or more public identifiers, which makes the dataset easily extensible. Paper nodes can be linked via arXiv ID and S2\footnote{\url{https://www.semanticscholar.org}} ID. Author nodes can be linked via S2 ID, and code repositories via GitHub's `\emph{user/repository}' identifier format (as used by the official GitHub API). These identifiers facilitate the integration of additional features from external sources through simple join operations.

\subsection{Data Collection}

We give a detailed description of the dataset collection process. We provide entity counts corresponding to a snapshot taken in June 2026, in order to give a sense of the orders of magnitude of the graph.

\textbf{GitHub Data} We collect all GitHub data using the GraphQL\footnote{\url{https://docs.github.com/en/graphql}} API. %
Initially, we search for repositories that mention the keyword “\emph{arxiv}” (case insensitive) in their README file, corresponding to approximately 444,000 repositories. For each repository, we then retrieve its engagement metrics, i.e. the number of stars, forks, issues, and pull requests, along with their associated timestamps.

To extract references to arXiv papers, we parse the main branch's README contents using regular expressions. We capture mentions in various formats such as URLs, e.g. `\emph{arxiv.org/abs/1234.5678}', or as in-text mentions, e.g. `\emph{arXiv:1234.5678}'. Since README files may evolve over time, we traverse each repository’s commit history to capture additions and deletions of arXiv links. To optimize this traversal and avoid impractical runtimes, we implement a binary search strategy. An interval between two commits is examined only if the set of arXiv links differs between the two endpoints. The output is a chronological list of tuples containing a timestamp and the corresponding set of arXiv IDs referenced at that time.

Collecting a complete snapshot from scratch takes around one month of querying and is primarily constrained by GitHub's API rate limits.

\textbf{Academic Graph} The core \emph{paper} entities are AI arXiv papers, obtained from the arXiv Open Archive Initiative (OAI) dataset\footnote{\url{https://info.arxiv.org/help/oa/index.html}}. A paper is included if any of its categories is in `\emph{cs.AI}',`\emph{cs.LG}',`\emph{cs.CL}' or `\emph{cs.CV}'. This provides approximately 558,000 papers along with their associated metadata (e.g. title, abstract, category...). We construct a citation graph centered on these papers, using the Semantic Scholar Open Research Corpus\footnote{\url{https://api.semanticscholar.org/api-docs/datasets}} \cite{Kinney2023TheSS}, by aggregating their incoming citations (including non-arXiv papers) and timestamping each edge using the citing paper’s publication date.

The set of \emph{author} entities is built from the union of all authors listed on arXiv AI publications, resulting in over 820,000 unique authors. We use Semantic Scholar’s Author Corpus ID to disambiguate between different name variations (e.g., “John Doe” vs. “Doe, J.”) \cite{s2and}. For each author, we retrieve their complete publication record (also including non-arXiv papers) in order to compute traditional academic metrics ($h$-index, $i$-10 index, total citation count..). This publication set spans approximately 11 million papers.

\subsection{Data Processing and Manipulation}

After data collection, we preprocess and merge the two data sources into a unified dataset.

\textbf{Publication Dates} ArXiv papers always have a precise publication date corresponding to their upload timestamp. In contrast, some papers outside of arXiv are sometimes only associated with a publication year (e.g., 2020). This mismatch in date formats is difficult to resolve, as the timing of citations can carry significant predictive value (a citation received within the first week after publication is very different from one received after six months). To mitigate this issue, we filter out citations with imprecise timestamps in our experiments. However, end users are free to process such data according to their own preferences or use cases.

\textbf{Merging the datasets} Given a specific snapshot date, we filter the entities, edges and features based on their timestamp. For academic entities, this involves selecting papers by publication date (which prunes the citation network) and then recovering the corresponding author features. For GitHub entities, we filter repositories on their creation date, recover engagement metrics and arXiv links from their README file at that time. Using arXiv IDs, we then perform a left join of papers with their associated authors and repositories, and group those features into lists. For each paper, this yields a list of author features and a list of repository features.

\section{Experiments}

We now present three experiments to illustrate key properties of the GitScholar dataset.

\subsection{Top Paper Prediction}

The goal of this experiment is to evaluate whether GitHub features provide useful information for identifying papers that will become influential over a relatively short-term horizon. To this end, we formulate paper impact as a fixed-horizon prediction task.

At the time of publication, a paper's future impact cannot be directly observed. We therefore define impact using a retrospective citation-based measure: for each paper, we look at the number of citations it has received exactly one year after publication. The one year horizon is fixed across all papers and is chosen to provide sufficient time for citations to accumulate, on average. This yields a static target that can be determined retrospectively for every paper in the dataset. This setup is fundamentally different from predicting a paper’s citation count at an arbitrary point in time, i.e. as an inductive rule.

To investigate the contribution of GitHub information, we train graph neural networks using different subsets of entities and features. We then compare their predictive performance across these configurations. In particular, this allows us to isolate the contribution of GitHub information and assess whether it provides additional predictive value beyond traditional academic features.

\textbf{Prediction Task} Suppose that we are given a set of $N$ papers. Each paper is connected to its existing neighbors with the various node and edge types outlined in the previous section. In addition, every paper has a feature time series vector $\bar{X}_t$, and a citation count series $Y_t$. Time $t$ is defined relative to each paper's publication date $t=0$. 

For each paper, our objective is to predict $Y_{1y}$, i.e. its citation count exactly one year after publication. We make predictions at weekly intervals after publication, using the corresponding feature vectors $\bar{X}_0, \bar{X}_{1w}, \dots, \bar{X}_{52w} = \bar{X}_{1y}$ and the neighboring subgraph at that time. We then measure a model's performance for each time lag $t$ by considering the predictions made for all papers at that relative time. In other words, we retrospectively align predictions made at the same paper ages.

\textbf{Evaluation Metric} For each time lag $t$, we rank the predicted citation counts and compare the top-K predicted papers with the top-K papers based on ground-truth citation counts. This yields a precision score that is easily interpretable, i.e. a proportion of papers correctly identified as a top paper at time lag $t$.  Note that the model we train is not optimized directly for this ranking, but for MSE of $\log$(1+citation count). 

\textbf{Entity and Feature Sets} We evaluate four entity sets, each used to train a model, and compare performance on the prediction task. Every value is read as of the exact day the prediction is made, signed-log transformed and standardized.
\begin{enumerate}
  \item \textit{Citation Baseline}: We use as feature the current citation count $Y_t$ and the time since publication $t$. (2 features)
  \item \textit{Citation+Author}: Extends the baseline by adding Author entities and the corresponding authorship edges. An author is described by the number of papers they had published by $t$ (1 feature), and their graph embedding is computed by sampling neighboring papers,  each described by its age, its citation count at $t$, and its number of authors (3 features per paper).
  \item \textit{Citation+Git}: Extends the baseline by adding GitHub Repository entities and their arXiv link edges. A repository is described by its number of stars, forks, issues and pull requests at $t$, together with the stars and forks it gained since it first linked the paper (6 features). The repository's graph embedding is computed through mentioned papers, each described by its age, its citation count at $t$, how many repositories mention it, and the star count and star gain of its best-starred repository (5 features per paper). The paper being scored also carries its own count of live repositories, giving it 3 features.
  \item \textit{Citation+Git+Author}: Combines all of the above nodes, edges and features. The paper being scored carries 4 features, authors 1, repositories 6, and each sampled neighboring paper 6.
\end{enumerate}

\textbf{Experimental Setup} We consider papers from mainstream AI topics, i.e. whose arXiv categories contain any of `\emph{cs.AI}', `\emph{cs.LG}', `\emph{cs.CL}' or `\emph{cs.CV}'. The training set contains papers published between 2021-03-21 and 2023-03-20 ($N=101,428$), and the validation set contains papers from 2023-03-21 to 2023-06-20 ($N=18,115$). The test set contains the papers published between 2024-06-18 and 2025-06-17 ($N = 94,525$). To avoid data leakage, it is important that all the papers from the validation set are published after the cutoff from the training set. The one year difference between the end of the validation set and start of the test set ensures that all the data required for training is available at the time of the first prediction in the test set. For precision measurement, we set $K=1000$ most impactful papers to detect.

We implement a simple 2-layer Graph Neural Network that predicts one-year citation counts as target variable. We train each configuration until convergence on the validation set, with AdamW \cite{adamw} and a batch size of 1024. Final predictions are made by averaging predictions from 3 random samples of a paper's neighborhood. We set the hidden dimension $D = 128$, learning rate $\eta= 10^{-3}$ and weight decay $\lambda=10^{-5}$. We performed a simple hyperparameter search over these parameters which did not move the MSE more than a standard deviation from neighborhood sampling. This suggests that the model is not capacity limited and is able to correctly extract the signal from the data. We also experimented using a LightGBM \cite{ke2017lightgbm} model, the comparison is in the appendix.

\begin{figure}[t]
	\includegraphics[width=1\columnwidth]{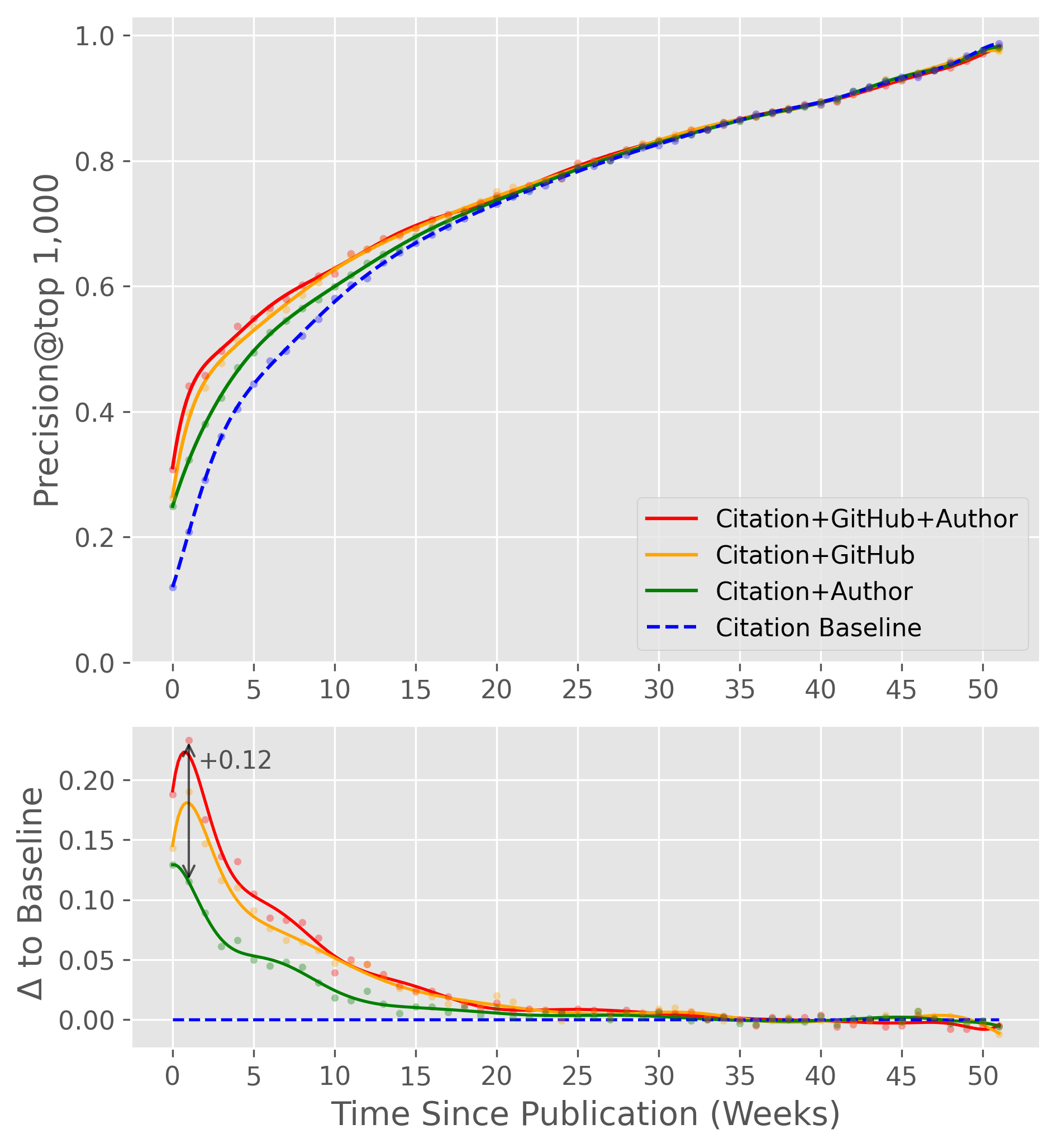} 
	\caption{Precision as a function of time since publication for models trained on different feature sets.}
	\label{fig:multi_plot2}
\end{figure}

\textbf{Results} Figure \ref{fig:multi_plot2} presents the results. Several key patterns can be noted:

\begin{itemize}
	\item Across all models, performance improvements are the highest shortly after publication, and decay roughly in a power law fashion. In the absence of citation data, a paper's future impact can be inferred either from the authors' academic reputation, or from early reactions on GitHub. As time passes, citations accumulate and become the dominant indicator of future success, outweighing author-level and GitHub features. 
	\item GitHub features (yellow curve) peak improvement one to two weeks after publication, and consistently surpass the author-only model (green curve). This is a striking result: a paper's early reception on GitHub provides signal as strong as, or even stronger than, the academic reputation of its authors. From the perspective of academic convention, it is surprising that a community-driven platform could match traditional indicators when predicting future academic impact.
	\item Incorporating GitHub features with the academic model outperforms using either model alone. At its peak, the combined approach yields an absolute 12\% improvement in precision over a strong academic baseline with full author and citation features.
\end{itemize}

\subsection{Mention Coverage}

We now present an experiment to estimate the GitHub \emph{mention coverage}, defined as the fraction of arXiv papers that are cited in at least one repository. This is a crucial metric to understand since the overall improvements provided by GitHub features are inherently limited by the proportion of papers that they cover, in a similar spirit to Amdahl's law.

The experiment is conducted as follows: considering a large subset of arXiv papers, i.e. those published between 2024-06-18 and 2025-06-17 (around 258,000 papers), we recover each paper's citation count exactly one year after its publication date. We then rank the papers according to this count, and create subsets of the ranking at various thresholds (e.g. the top 10\%, 5\%, 1\%, etc.). For each subset, we report the fraction of papers that were mentioned on GitHub within one year after their publication.

The purpose of such thresholding is to measure the coverage of papers at various levels of academic impact, roughly approximated by their one-year citation growth.

\begin{figure}[t]
	\centering
	\includegraphics[width=\columnwidth]{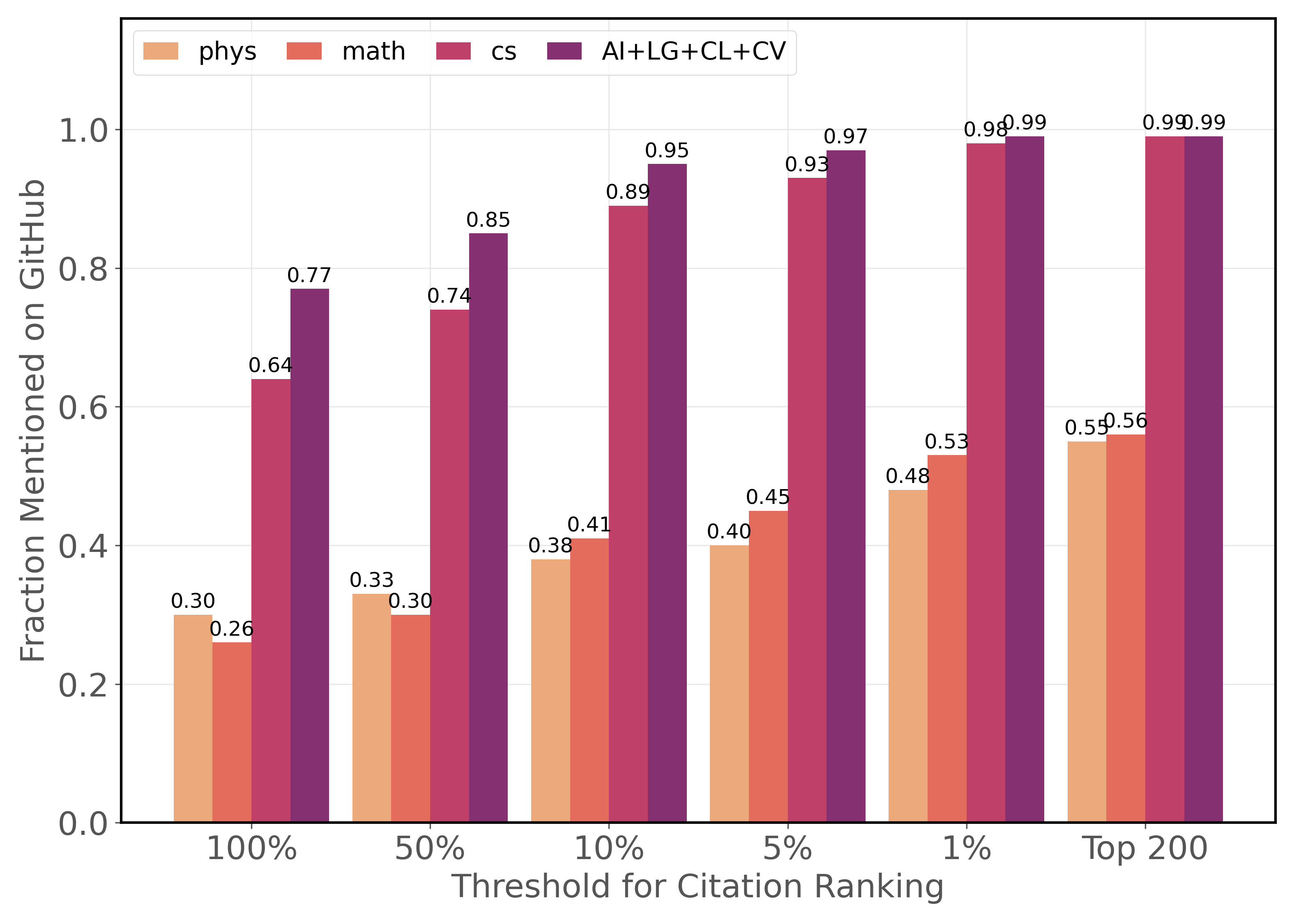}
	\caption{Fraction of GitHub-mentioned arXiv papers, per research domain and per threshold of one-year citation count ranking.}
	\label{fig:bar}
\end{figure}

Furthermore, we group papers according to their high-level research domains using their arXiv categories. The groups are: 1) Physics: all physics-related categories, e.g. `\emph{astro-ph}', `\emph{cond-mat}', `\emph{hep-th}', etc.. 2) Mathematics: all subcategories under `\emph{math.*}', 3) Computer Science: all subcategories under `\emph{cs.*}', and 4) Mainstream AI topics: all papers in any of `\emph{cs.AI}', `\emph{cs.LG}', `\emph{cs.CL}' or `\emph{cs.CV}'. The results are given in Figure \ref{fig:bar}.

 We can observe the following:

1) GitHub coverage differs significantly across research domains. Computer Science papers are much more represented than those in Physics or Mathematics, which is understandable since GitHub is a code sharing platform. More notably, mainstream AI categories are among the most highly covered subset of Computer Science topics. This confirms the intuition that GitHub is already a widely used platform specifically in AI research. For other disciplines, our dataset could potentially provide extra signal, but most likely with limitations due to the lower coverage (and potentially different community behavior).

2) Across domains, impactful papers (as estimated by early citation growth) are more likely to be represented on GitHub. In AI especially, high-impact papers are rarely absent, with GitHub achieving near-perfect coverage for the top 1\% papers. By filtering AI papers that are not present on GitHub, one removes 23\% of AI papers but almost none that are high impact. This suggests that both the presence and absence of an AI paper on GitHub can be interpreted as meaningful signal (unlike for Physics where almost half of the impactful papers are simply missing from GitHub).

\subsection{Comparison with Twitter}

Over the past few years, $\mathbb{X}$ (formerly Twitter) has been a popular platform for disseminating and discovering scientific papers. Data from $\mathbb{X}$ is not freely available, but we explore how engagement signals from GitHub compare to those derived from a previous snapshot of Twitter. More specifically, we examine two metrics: (1) the Spearman correlation between a paper's total number of likes across tweets that mention it and its citation count one year later, and (2) the Spearman correlation between a paper's total number of GitHub stars across repositories referencing it and its citation count one year later.

We use the TweetPap dataset \cite{jain2021}, which, to the best of our knowledge, is the most recent suitable dataset publicly available. It includes all tweets containing the keyword ``\emph{arXiv}'' between 2010 and 2019, or over 330,000 tweets covering more than 125,000 arXiv papers.

We observe that a substantial fraction of the data from that period consists of bot-generated content. For example, we find that 9 out of the 10 most active accounts are bots and are responsible for over $191,000$ tweets, or around 58\% of the dataset. These bots often tweet indiscriminately about large volumes of arXiv papers, sometimes every paper in a given category (e.g. '\emph{@arxiv\_cs\_cl}'), without generating meaningful engagement.

To make our correlation estimates more representative, we manually identified and removed the 30 most active bot accounts. The resulting filtered dataset contains 70,000 tweets referencing 52,000 papers (41\% of the original count). While some bot activity may remain, we believe this filtering significantly reduces noise.

For GitHub data, we extract a snapshot from 2019-11-01 (estimated scrape date from TweetPap), which yields around 42,000 repositories referencing around 33,000 arXiv papers. The two Spearman rank correlations are given in Table \ref{table:corr}.

\begin{table}[h]
	\setlength{\tabcolsep}{3pt} 
	\begin{center}
		\begin{tabular}{lcl}
			\toprule
			\multicolumn{1}{l}{Feature} & {Spearman Correlation} & $p$-value   \\
			\midrule
			Likes                       & -0.103                 & $<10^{-10}$ \\
			Stars                       & \hspace{0.35em}0.259   & $<10^{-10}$ \\
			\bottomrule
		\end{tabular}
	\end{center}
	\caption{Spearman rank correlation between features and future citation count.}
	\label{table:corr}
\end{table}

\begin{figure}[t]
	\centering
	\includegraphics[width=0.93\columnwidth]{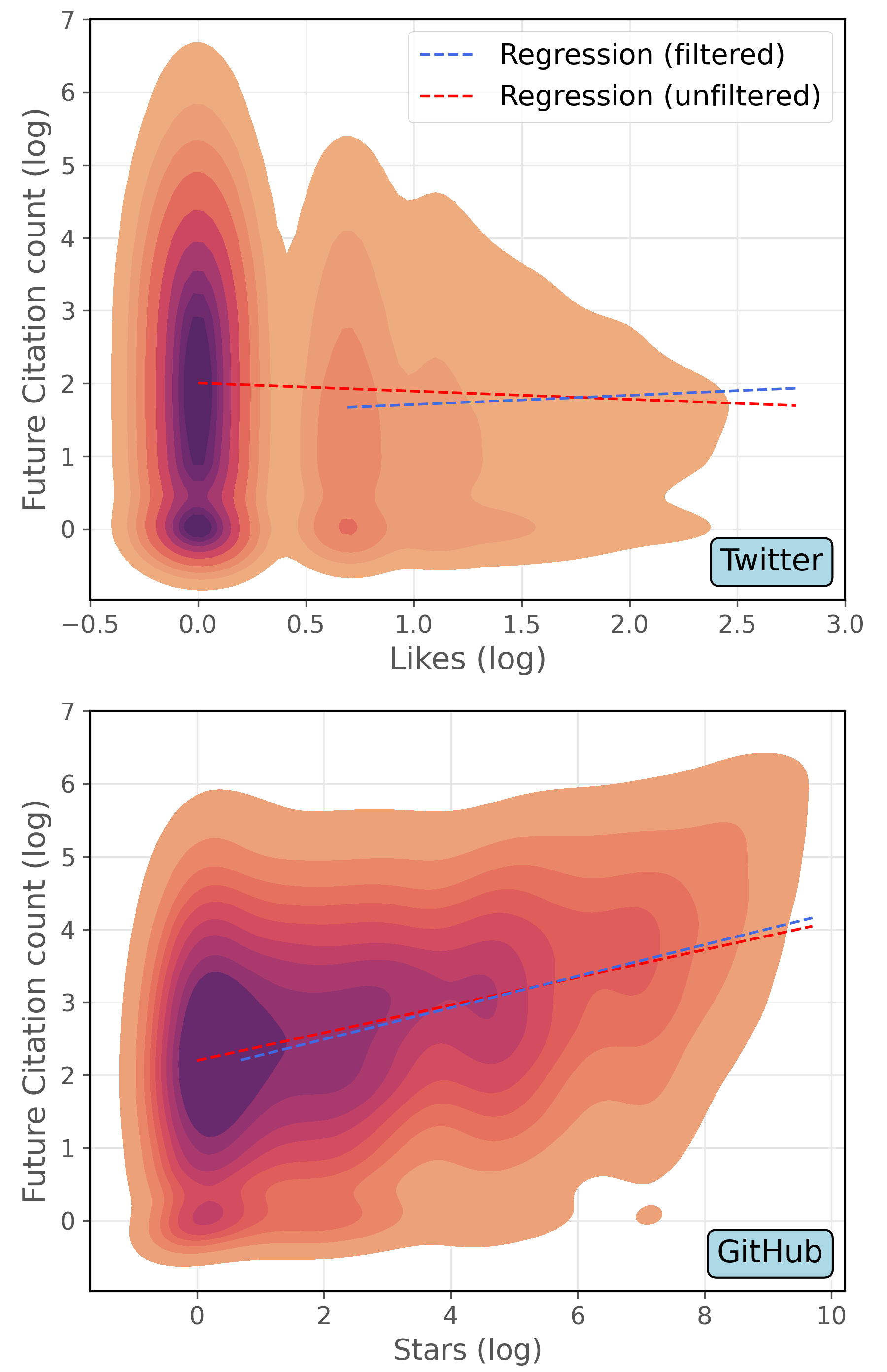}
	\caption{Correlation density plots between Twitter likes (top) or GitHub stars (bottom) with future citation count. The dashed lines are linear regressions fitted on raw (red) and filtered data (blue).}
	\label{fig:x_density}
\end{figure}

We observe a positive correlation for GitHub stars and surprisingly, a slight negative correlation for Twitter likes. Upon visual inspection of the density plots (Figure \ref{fig:x_density}), we find that the Twitter dataset includes a substantial number of ``contradictory samples'', i.e. papers that did not generate engagement but ultimately received many citations. These samples are disproportionately represented and skew the overall correlation downward. To address this, we repeat the experiment using only papers that received at least one like on Twitter. For comparison, we apply a similar filter to the GitHub dataset, keeping only papers that received at least one star. The updated results are shown in Table \ref{table:corr2}.

\begin{table}[h]
	\setlength{\tabcolsep}{3pt} 
	\begin{center}
		\begin{tabular}{lcl}
			\toprule
			\multicolumn{1}{l}{Feature} & {Spearman Correlation} & $p$-value   \\
			\midrule
			Likes (filtered)            & 0.035                  & $<10^{-4}$ \\
			Stars (filtered)            & 0.272                  & $<10^{-10}$ \\
			\bottomrule
		\end{tabular}
	\end{center}
	\caption{Spearman rank correlation between features and future citation count, on filtered data.}
	\label{table:corr2}
\end{table}

After filtering, we observe a small positive correlation in the Twitter data, and a slight increase in the correlation for GitHub. Interestingly, this filtering substantially impacts the Twitter correlation, while the effect on GitHub is minimal. We can quantify this difference using the absolute relative change in correlation, defined as:

\[ \Delta= \left|\frac{\rho_f - \rho_r}{\rho_r}\right| * 100\]

where $\rho_r$ and $\rho_f$ are the correlation coefficients for the raw and filtered data, respectively.

For GitHub, we find $\Delta_G = 5\%$, while for Twitter we observe a much larger change of $\Delta_T = 131\%$. This significant shift for Twitter after filtering suggests a sharp discontinuity in the sample distribution, which is clearly visible on the density plot. When a paper receives zero likes, this lack of engagement does not reliably indicate a lack of future academic impact, due to the high proportion of contradictory samples\footnote{It remains unclear whether this effect is due to residual bot activity or to conflicting patterns of user engagement, e.g. influential academics who are not particularly popular on Twitter.}. For papers that do generate some engagement, the correlation with citations remains weak overall. 

In contrast, GitHub shows a clear positive correlation, with minimal change after filtering. This suggests that samples are consistently distributed across the popularity spectrum, and that both low and high levels of GitHub activity are on average reliable indicators of lower and higher future academic impact, respectively.

Since the experiment is based on data from 2019, we cannot rigorously draw conclusions about the present-day comparison. However, we can look at GitHub data only, and see how the correlation has evolved since. Using all the AI arXiv papers published between 2024-06-18 and 2025-06-17 (around 95,000 papers), we compute the Spearman rank correlation between their citation count at one year of age, and the sum of their stars at cutoff 2025-06-17. We find a correlation of 0.374 ($p$-value $< 10^{-10}$), an approximate 44\% increase in the past 6 years. This suggests that GitHub engagement has become an increasingly reliable proxy for future academic impact of AI papers.

\section{Dataset Release and License}

GitScholar is available at \url{https://huggingface.co/datasets/huawei-csl/GitScholar}. It combines public repository metadata collected from GitHub and academic paper metadata from the Open Archives Initiatives (OAI) and the Semantic Scholar Open Research Corpus \cite{Kinney2023TheSS}. All GitHub data was collected via the official GitHub GraphQL API and includes only public metadata, in accordance with GitHub’s API Terms of Use. No personal data, user-generated content, or private repository information is included.
It is released under the permissive ODC-By license, allowing for reuse and modification with appropriate attribution.
\section{Related Work}

\textbf{Academic Impact Prediction}
Research impact prediction is a well-studied problem within the fields of scientometrics and bibliometrics \cite{info8030073}, with two common high-level approaches. The first approach focuses primarily on correctly modeling citation trajectory. For example \citet{wang2013quantifying} model citations 30+ years after publication by fitting per-paper parameters on the first 10 years of citation data; \citet{cao2016data} predict 5-year citation counts by matching papers with similar citation patterns in their first 3 years. While effective, these methods rely on extended initial citation windows, which are impractical in fast-moving fields like AI. The second approach focuses on identifying and combining various academic impact indicators that provide useful signal, such as author reputation (e.g. $h$-index), journal prestige (e.g. impact factor), and early citation counts. For example, \citet{Stegehuis_2015} apply quantile regression using journal impact factor and citation data from the first year post-publication; \citet{Yu2014Citation} use stepwise regression with similar inputs, including author-level features. In our setting, i.e. when assessing recent arXiv preprints, only author-level features and early citation signals (e.g. from the first weeks) are consistently available, which we use as baseline in our experiments.

\textbf{Altmetrics and Social Media}
Since traditional academic indicators usually require time to accumulate, alternative earlier signals have been explored, commonly referred to as `altmetrics'. These include social media reactions across platforms such as Twitter, Reddit, Facebook, or LinkedIn \cite{thelwall2013altmetrics}. While altmetrics have been shown to correlate positively with later citation counts, the relationship is generally weak, suggesting that they tend to reflect social visibility rather than scientific impact \cite{costas2015}. Our experiments support these findings, and further show that engagement on a technical platform like GitHub exhibits much stronger and more consistent correlation with future academic impact.

\textbf{GitHub in the AI Ecosystem} GitHub plays a central role in the AI research community, where openness, collaboration and reproducibility are key objectives. Reflecting this close relationship, \citet{gonzalez2020} showed that AI-related repositories exhibit markedly different engagement and collaboration patterns compared to typical GitHub software projects. Furthermore, \citet{KANG2023103477} found that, for papers listed on \emph{Papers With Code}, the presence of a linked repository was associated with a significantly higher citation count. \citet{bhattarai2022} analyzed repositories whose URLs were explicitly cited in papers from eight top computer science conferences and found that engagement with these repositories could help predict highly cited papers. However, their analysis is limited to direct one-to-one links from papers to repositories (presumably owned by the authors), and focused exclusively on peer-reviewed conference publications. While our results are consistent with their findings, we aim to provide the first full-scale, comprehensive analysis of GitHub reactions as an early signal specifically for predicting future AI research impact. To the best of our knowledge, we are also the first to release such a dataset.
\section{Conclusion}

We present GitScholar, a large-scale dataset that connects the early reception of AI papers on GitHub with their subsequent academic impact. Our experiments show that GitHub reactions offer both an effective and early signal for predicting future academic success. In particular, incorporating GitHub reactions improves citation prediction precision in the initial weeks after publication, achieving a 12\% absolute gain over a strong academic baseline. Remarkably, these reactions provide a signal comparable in strength to traditional academic indicators, such as author reputation. Furthermore, we show that GitHub offers near-complete coverage of high-impact AI papers, and that engagement on the platform is consistently correlated with future academic success. Finally, we find that the quality of GitHub signal has increased in recent years, reflecting the platform's growing adoption within the AI research community. Based on these findings, we believe that GitHub offers a strong foundation for bringing clarity at scale to the current AI research landscape. To encourage further research in this direction, we release GitScholar as FAIR-compliant, continually updated and openly licensed dataset.

\section*{Impact Statement}

This work aims to improve academic impact prediction by incorporating GitHub engagement metrics alongside traditional academic features. The authors acknowledge several limitations to this approach. Firstly, citation count as a metric for a paper's impact is an approximation, and is used because it is measurable. Secondly, GitHub engagement is less regulated than paper citations, making these metrics vulnerable to manipulation through fake engagement. Lastly, automated impact prediction methods can influence how publications are assessed, and their fairness should be further investigated. The purpose of such methods should be to learn a probabilistic prior about papers given their public reactions. This allows for a large scale approximate ranking of papers, but not an accurate assessment of any individual paper.

\bibliography{bib}

\clearpage
\raggedbottom %
\appendix

\begin{center}
	{\Large\bfseries Appendix}
\end{center}
\vspace{-0.5em} 

\section{Comparison between GNN and LightGBM}

\begin{figure}[H]
	\centering
	\includegraphics[width=0.93\columnwidth]{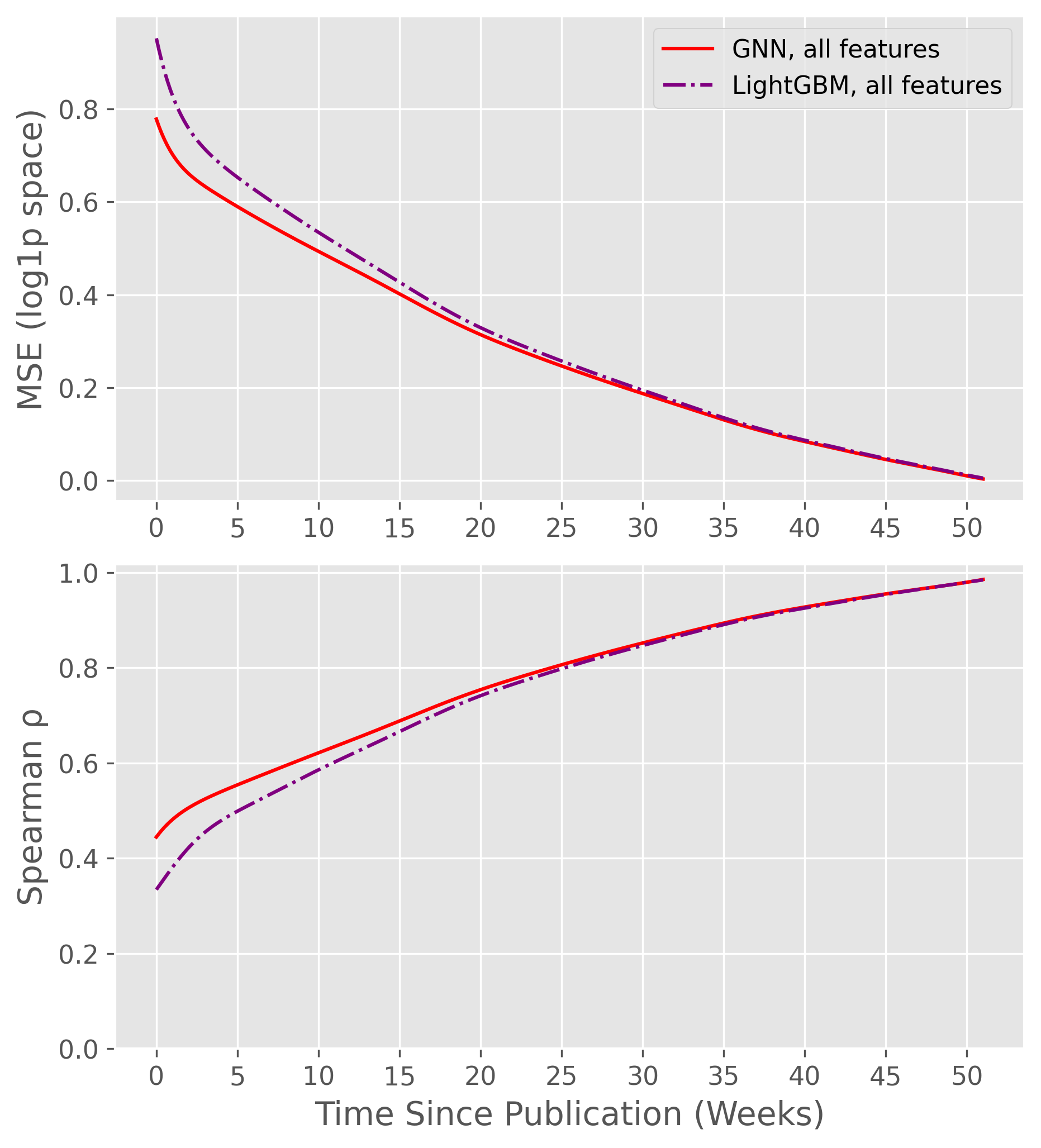}
	\caption{MSE and Spearman rank comparison between GNN and LightGBM.}
	\label{fig:mse_gbm}
\end{figure}

\begin{figure}[H]
	\centering
	\includegraphics[width=0.93\columnwidth]{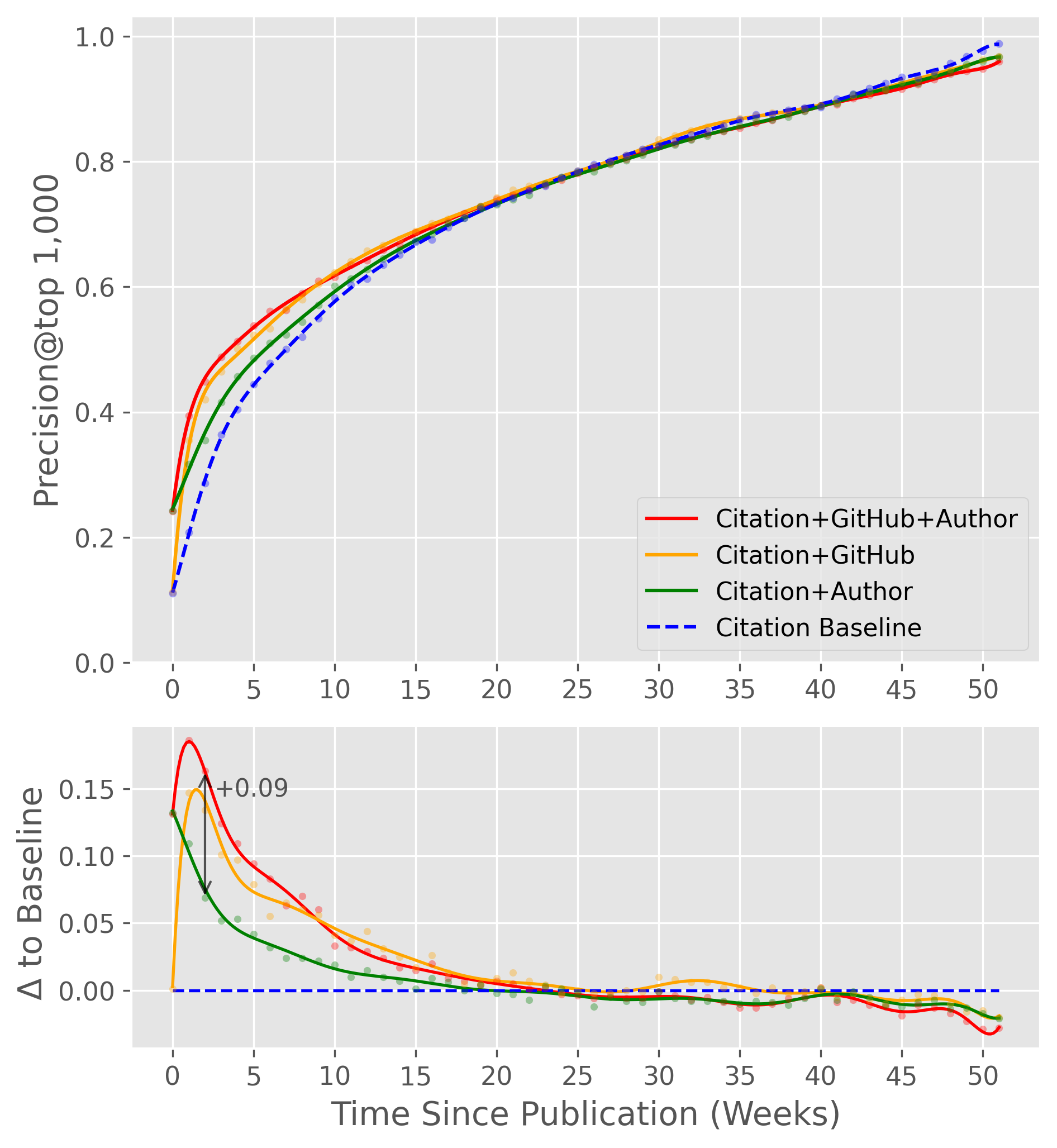}
	\caption{Prediction experiment using LightGBM instead of the GNN.}
	\label{fig:pred_gbm}
\end{figure}

We compare the LightGBM model with the GNN using the same input data. The GNN is a more expressive model, since it leverages relational signal between entities. It achieves a lower MSE and better performance in ranking papers in Figure \ref{fig:mse_gbm}. In Figure \ref{fig:pred_gbm}, we reproduce the same prediction experiment using LightGBM. The prediction patterns are very similar to those in Figure \ref{fig:multi_plot2}. This suggests that these patterns are primarily driven by the underlying data rather than being specific to the choice of model architecture.

\end{document}